\documentclass[reprint,showpacs,dcolumn,preprintnumbers,amsmath,amssymb,prc,floatfix]{revtex4-1}
\usepackage{float}
\usepackage{color}
\usepackage{graphicx}
\usepackage{dcolumn}
\usepackage{bm}
\usepackage{xcolor}
\usepackage{times}
\usepackage{multirow}
\usepackage{threeparttable}
\usepackage{adjustbox}
\usepackage{longtable}
\usepackage{comment}
\usepackage{natbib}
\usepackage{array}
\usepackage[colorlinks,citecolor=blue,linkcolor=red,anchorcolor=blue,filecolor=blue,urlcolor=blue]{hyperref}

\begin{document}
	
	\title{Thermal impacts on the properties of nuclear matter and young neutron star} 
	\author{Ankit Kumar$^{1,2}$}
	\email{ankit.k@iopb.res.in}
	\author{H. C. Das$^{1,2}$}
	\email{harish.d@iopb.res.in}
	\author{M. Bhuyan$^{3,4}$}
	\email{bunuphy@um.edu.my}
	\author{S. K. Patra$^{1,2}$}
	\email{patra@iopb.res.in}
	\affiliation{\it $^{1}$Institute of Physics, Sachivalya Marg, Bhubaneswar 751005, India}
	\affiliation{\it $^{2}$Homi Bhabha National Institute, Training School Complex, Anushakti Nagar, Mumbai 400094, India}
	\affiliation{$^3$Center for Theoretical and Computational Physics, Department of Physics, Faculty of Science, University of Malaya, Kuala Lumpur 50603, Malaysia}
	\affiliation{$^4$Institute of Research and Development, Duy Tan University, Da Nang 550000, Vietnam}
	
	\date{\today}
	
	\begin{abstract}
		We present a methodical study of the thermal and nuclear properties for the hot nuclear matter using relativistic-mean field theory. We examine the effects of temperature on the binding energy, pressure, thermal index, symmetry energy, and its derivative for the symmetric nuclear matter using temperature-dependent relativistic mean-field formalism for the well-known G2$^{*}$ and recently developed IOPB-I parameter sets. The critical temperature for the liquid-gas phase transition in an asymmetric nuclear matter system has also been calculated and collated with the experimentally available data. We investigate the approach of the thermal index as a function of nucleon density in the wake of relativistic and non-relativistic formalism. The computation of neutrino emissivity through the direct Urca process for the supernovae remnants has also been performed, which manifests some exciting results about the thermal stabilization and evolution of the newly born proto-neutron star. The central temperature and the maximum mass of the proto-neutron star have also been calculated for different entropy values.
	\end{abstract}
	
	\maketitle
	
	\section{Motivation}\label{intro}
	The neutron stars are the products of supernova explosion or merger remnants of two compact stars and well-known ideal laboratory to explore dense matter objects' formulated theories. The gravitational collapse of massive stars in type-II supernova happens due to the mass accretion process, which results in a far more fascinating and brighter explosion than an entire galaxy \cite{Bethe_1990,Shapiro_1983,Burgio_2016}. The mass accretion process coerced the stars to create a Fe core, and the core bounce occurs when the Fe core mass reaches the Chandrasekhar mass limit. This outburst of a massive star with $Fe$- cores are generally categorized into two types of explosions- prompt and delayed blast. The equation of state (EoS) of nuclear matter is the most crucial element to understand the mechanism of both types of explosions \cite{Bethe_1990,Shen_1998}. \\ 
	A very hot lepton-rich object, along with trapped neutrinos, known as proto-neutron star, is formed just after the supernovae explosion. To explore the mechanical, nuclear, and thermal properties of this newly formed proto-neutron star (PNS), we need a detailed understanding of the EoS of the nuclear matter for a wider range of density and temperature \cite{Kumar_2020}. The subsequent evolution and thermal stabilization of the PNS are further controlled by the neutrino reaction rates and trapped neutrinos' emission. The high photon and neutrino emissivities are mainly amenable for the agile cooling of young PNS. The larger part of the process is dominated by the neutrino emissivity \cite{BOGUTA_1981}. The dynamics of the collapsed core and the supernovae explosion are also significantly controlled by the neutrino emissivity \cite{Burrows_2000, Barbara_2003}. The cooling rate of the hot dense astrophysical object is enhanced by the emission of trillions of neutrinos, which is generally controlled by two channels, direct and modified Urca process \cite{TSANG_2019,Gusakov_2002}. In the present work, we explicitly explore the effects of the direct Urca process on the cooling mechanism of newly born PNS with the help of the EoS. Due to the rapidity of core-collapse supernovae explosion's evolution, the matter is incapable of attaining $\beta$-equilibrium condition for a short period, so the EoS for a dense nuclear matter can be used to paraphrase the dynamics of core-collapse event \cite{Anthony_2005,Schneider_2019}. \\
	At present, the relativistic mean-field (RMF) theory is the most essential and satisfactory phenomenological model to understand the statics and dynamics of finite nuclei and astrophysical objects. Therefore, here we provide an approach to explore the nuclear and thermal properties of hot nuclear matter and newly born proto-neutron star with RMF temperature-dependent EoS being the main ingredient. We extend the RMF formalism to finite temperature and examine various nuclear and thermal properties like phase transition temperature, binding energy, symmetry energy, and its derivatives for infinite nuclear matter. The dynamics of exotic nuclei and heavy-ion collision, obtained through experimental facilities such as  GANIL-SPIRAL2 \cite{GANIL} and FRIB-USA \cite{FRIB_2007},  can also be appropriately interpreted with the help of finite temperature EoS. Finite temperature EoS can be calibrated through the outputs of the heavy-ion collision experiments in terrestrial laboratories, which can produce hot and dense nuclear matter and provide information about the nuclear matter's thermodynamical properties \cite{Ono_2003,Iglio_2006}. Various nuclear and thermal properties such as pressure (P), incompressibility, the thermal index ($\Gamma_{th}$), free symmetry energy (F$_{sym}$) and its derivatives, i.e., slope (L$_{sym}$), curvature (K$_{sym}$) and skewness (Q$_{sym}$) parameter are strongly correlated with neutron drip-line, neutron skin thickness in super-heavy nuclei and gravitational binding energy of neutron star \cite{TSANG_2019,Typel_2018}. Hence, the density dependence of the finite temperature EoS can be constrained with experimentally available data of all these nuclear and thermal properties. We studied these properties using the popular G2$^*$ \cite{Sulaksono_2005, PhysRevC.74.045806} and recently developed IOPB-I \cite{Bharat_2018} parameter sets. We did a comparative analysis, i.e., how differently these nuclear and thermal properties of the symmetric nuclear matter vary for the prescribed G2$^*$ and IOPB-I forces. The mass-radius (M-R) profile of the PNS for a certain temperature range has also been studied by imposing the Tolman-Oppenheimer-Volkoff (TOV) equations on the EoS of both the parameter sets.\\
	The paper has been stated as: The theoretical framework used in the present work is discussed in the sec. \ref{Formalism}. All the results are presented in sec. \ref{result} and has been discussed explicitly. Finally, concluding remarks are offered in sec. \ref{conc}.\\
	
	\section{Theoretical Formalism}\label{Formalism}
	\subsection{RMF Model for Finite Temperature}\label{RMF}
	\noindent
	The basic idea to construct a model based on relativistic quantum field theory to explore the properties of infinite nuclear matter was first proposed by J. D. Walecka in 1974. It is a simple model where nucleons are described as Dirac spinors which can interact through the exchange of isoscalar scalar, and isoscalar vector mesons, namely $\sigma$ and $\omega$ \cite{WALECKA_1974,GAMBHIR_1990}. This RMF theory has many advantages over the non-relativistic model. It resolves the coester band problem and incorporates the spin-orbit interaction, which predicts many nuclear properties up to a surpassing accuracy \cite{DELFINO_2005}. Afterward, isovector vector mesons ($\rho$ and $\delta$) and several terms with self and cross-coupling of mesons up to 4$^{th}$ order are taken into consideration for a better enucleation of many experimental results and the Lagrangian density along with most of the desired interactions reads as \cite{Shailesh_2014,Kumar_2020}:
	\begin{widetext}
		\begin{eqnarray}
		{\cal L} & = &  \sum_{\alpha=p,n} \bar\psi_{\alpha}
		\Bigg\{\gamma_{\mu}(i\partial^{\mu}-g_{\omega}\omega^{\mu}-\frac{1}{2}g_{\rho}\vec{\tau}_{\alpha}\!\cdot\!\vec{\rho}^{\,\mu})
		-(M-g_{\sigma}\sigma)\Bigg\} \psi_{\alpha}
		+\frac{1}{2}\Bigg(
		\partial^{\mu}\sigma\,\partial_{\mu}\sigma
		-m_{\sigma}^{2}\sigma^2\Bigg) -g_{\sigma}\frac{m_{\sigma}^2}{M}\Bigg(\frac{\kappa_3}{3!}
		+ \frac{\kappa_4}{4!}\frac{g_{\sigma}}{M}\sigma\Bigg)
		\sigma^3
		\nonumber \\[3mm]
		& & \null 
		+\Bigg(\frac{1}{2}m_{\omega}^{2}\omega^{\mu}\omega_{\mu}
		-\frac{1}{4}F^{\mu\nu}F_{\mu\nu}\Bigg)+\frac{\zeta_0}{4!}g_\omega^2
		(\omega^{\mu}\omega_{\mu})^2 +\frac{1}{2}\frac{g_{\sigma}\sigma}{M}\Bigg(\eta_1+
		\frac{\eta_2}{2} \frac{g_{\sigma}\sigma}{M}\Bigg)m_\omega^2\omega^{\mu}\omega_{\mu} +\Bigg(\frac{1}{2}m_{\rho}^{2}\rho^{\mu}\!\cdot\!\rho_{\mu}
		-\frac{1}{4}\vec R^{\mu\nu}\!\cdot\!\vec R_{\mu\nu}\Bigg)
		\nonumber
		\\[3mm]
		& &  \null 
		+\frac{1}{2}\eta_{\rho}\frac{m_{\rho}^2}{M}g_{\sigma}\sigma(\vec\rho^{\,\mu}\!\cdot\!\vec\rho_{\mu})
		-\Lambda_{\omega}g_{\omega}^2g_{\rho}^2(\omega^{\mu}\omega_{\mu})
		(\vec\rho^{\,\mu}\!\cdot\!\vec\rho_{\mu}),
		\label{eq1}
		\end{eqnarray}
	\end{widetext}
	where M is the mass of the nucleons; $m_\sigma$, $m_\omega$ and $m_\rho$, $g_\sigma$, $g_\omega$ and $g_\rho$ are the masses and the coupling constants for the $\sigma$, $\omega$ and $\rho$ mesons respectively; $\kappa_3$ (or $\kappa_4$) and $\zeta_0$ defines the strength of self-interaction for the $\sigma$ and $\omega$ mesons respectively; and $\eta_1$, $\eta_2$, $\eta_\rho$ and $\Lambda_\omega$ stands for the strength of non-linear cross-coupling between mesons. The fields $F^{\mu\nu}$ and $\vec R^{\mu\nu}$ will be given by $F^{\mu\nu}$ = $\partial^\mu\omega^\nu-\partial^\nu\omega^\mu$ and $\vec R^{\mu\nu}$ = $\partial^\mu\vec\rho^{\,\nu}-\partial^\nu\vec\rho^{\,\mu}$.
	By imposing the mean-field approximation, where the ground state expectation value will be used in place of the original field operator and only the third isospin component for the neutron and proton remains, the Dirac equation can be derived as:
	\begin{eqnarray}
	\Bigg\{i\gamma_{\mu}\partial^{\mu}-g_{\omega}\gamma_{0}\omega-\frac{1}{2}g_{\rho}\gamma_{0}\tau_{3\alpha}\rho-M^\ast_\alpha\Bigg\} \psi_\alpha=0,
	\end{eqnarray}
	where $\alpha$ stands for proton (p) and neutron (n). $\tau_3$ is the isospin operator and $M^\ast_\alpha$ is the effective nucleon mass which by definition is given by:
	\begin{eqnarray}
	M_{p}^{\ast}&=&M-g_{\sigma}\sigma,  \\
	M_{n}^{\ast}&=&M-g_{\sigma}\sigma.
	\end{eqnarray}
	Further, the Euler-Lagrange equations were derived from the Lagrangian density using variational principle as \cite{Bharat_2018},
	\begin{widetext}
		\begin{eqnarray}
		m_{\sigma}^2\sigma & = & g_{\sigma} \sum_{\alpha=p,n}\langle\bar\psi_\alpha\gamma_0\psi_\alpha\rangle - \frac{m_\sigma^2g_\sigma}{M}\sigma^2\Bigg(\frac{\kappa_3}{2} + \frac{\kappa_4}{3!}\frac{g_\sigma\sigma}{M}\Bigg)+ \frac{g_\sigma}{2M}\Bigg(\eta_1 + \eta_2\frac{g_\sigma\sigma}{M}\Bigg){ m_\omega^2\omega^2} + \frac{\eta_{\rho}}{2M}{g_\sigma}{m_\rho^2 }{\rho^2}, \\
		m_\omega^2\omega & = & g_\omega \sum_{\alpha=p,n} \langle\bar\psi_\alpha\psi_\alpha\rangle - \Bigg(\eta_1 + \frac{\eta_2}{2}\frac{g_\sigma\sigma}{M}\Bigg)\frac{g_\sigma\sigma}{M}m_\omega^2\omega - \frac{1}{3!}\zeta_0g_{\omega}^2\omega^3 - 2\;\Lambda_{\omega} g_{\omega}^2 g_{\rho}^2 \rho^2 \omega, \\ 
		m_{\rho}^2 \rho & = &\frac{1}{2}g_{\rho} \sum_{\alpha=p,n} \langle\bar\psi_\alpha\tau_3\psi_\alpha\rangle - \eta_\rho \frac{g_\sigma\sigma}{M}m_{\rho}^2\rho - 2\;\Lambda_{\omega} g_\omega^2 g_\rho^2 \omega^2 \rho,
		\end{eqnarray}
		The ground state expectation values of the nucleon current i.e. $\langle\bar\psi_\alpha\psi_\alpha\rangle$, $\langle\bar\psi_\alpha\gamma_0\psi_\alpha\rangle$, $\langle\bar\psi_\alpha\tau_3\psi_\alpha\rangle$ and $\langle\bar\psi_\alpha\tau_3\gamma_0\psi_\alpha\rangle$ at finite temperature can be defined as \cite{Yang_2019,Wang_2000},
		\begin{eqnarray}
		n &=& \sum_{\alpha=p,n} \langle\bar\psi_\alpha\psi_\alpha\rangle  = n_{p}+n_{n} = \sum_{\alpha=p,n}\frac{2}{(2\pi)^{3}}\int d^{3}k\,[f_{\alpha}(\mu_{\alpha}^{\ast},T)-\bar f_{\alpha}(\mu_{\alpha}^{\ast},T)], \\ 
		n_s &=& \sum_{\alpha=p,n}\langle\bar\psi_\alpha\gamma_0\psi_\alpha\rangle = n_{sp}+n_{sn} = \sum_{\alpha=p,n} \frac{2}{(2\pi)^3}\int d^{3}k\, 
		\frac{M_{\alpha}^{\ast}} {(k^{2}_{\alpha}+M_{\alpha}^{\ast 2})^{\frac{1}{2}}}[f_{\alpha}(\mu_{\alpha}^{\ast},T)+\bar f_{\alpha}(\mu_{\alpha}^{\ast},T)], \\ 
		n_3 &=& \sum_{\alpha=p,n} \langle\bar\psi_\alpha\tau_3\psi_\alpha\rangle = n_p-n_n, \\ 
		n_{s3} &=& \sum_{\alpha=p,n} \langle\bar\psi_\alpha\tau_3\gamma_0\psi_\alpha\rangle = n_{sp}-n_{sn},
		\end{eqnarray}
	\end{widetext}
	Here, $T$ stands for the temperature and  $f_{\alpha}(\mu_{\alpha}^{\ast},T)$ and $\bar f_{\alpha}(\mu_{\alpha}^{\ast},T)$ are the familiar distribution function of nucleon and anti-nucleon at temperature T. $k_\alpha$ is the nucleon Fermi momentum and $\mu_\alpha^{\ast}$ is defined as the effective chemical potential of the nucleon, which for proton and neutron can be stated as, $\mu_{p}^{\ast}=\mu_{p}-g_{\omega}\omega-\frac{1}{2}g_{\rho}\rho$ and $\mu_{n}^{\ast}=\mu_{n}-g_{\omega}\omega+\frac{1}{2}g_{\rho}\rho$ respectively. The thermal Fermi distribution function for the nucleon and the anti-nucleon at temperature T can be written as \cite{Yang_2019},
	\begin{eqnarray}
	f_{\alpha}(\mu_{\alpha}^{\ast},T)&=&\frac{1}{e^{({\cal E}_{\alpha}^{\ast}-\mu_{\alpha}^{\ast})/k_BT}+1}, \\ \nonumber \\ 
	\bar{f_{\alpha}}(\mu_{\alpha}^{\ast},T)&=&\frac{1}{e^{({\cal E}_{\alpha}^{\ast}+\mu_{\alpha}^{\ast})/k_BT}+1},
	\end{eqnarray}
	with $k_B$ as the Boltzmann constant and ${\cal E_\alpha^\ast}$ being the effective energy of the nucleons in the mesonic field defined as ${\cal E_\alpha^\ast} = \sqrt{k_\alpha^2+M_\alpha^{\ast2}}.$
	\subsection{SYMMETRIC NUCLEAR MATTER}\label{SNM}
	We know that the total baryon number density of the nuclear matter is given by the number density of protons and neutrons. Such a condition leads us to the relation between the total Fermi momenta and the momentum of protons and neutrons, which is defined as \cite{Glendenning_1989},
	\begin{eqnarray}
	k_{p} &=& k_{F}(1-t)^{1/3},   \\ \nonumber \\
	k_{n} &=& k_{F}(1+t)^{1/3},
	\end{eqnarray}
	where $t$ is the asymmetry factor given by, $t=\frac{n_n-n_p}{n_n+n_p}$. Here, we mainly explore the nuclear and structural properties of symmetric nuclear matter (SNM), for which the value of t will be equal to zero, i.e., $n_{n}=n_{p}$. Now, the energy and the pressure pressure for the SNM can be obtained from the Lagrangian defined in the sec. \ref{RMF} with the help of energy-momentum tensor $T^{\mu\nu}$. So, the expression for the energy density and pressure of the SNM at finite temperature will be given by  \cite{Fetter_1971,Shailesh_2014},   
	\begin{widetext}
		\begin{eqnarray}
		E & = & \sum_{\alpha=p,n} \frac{2}{(2\pi)^{3}}\int d^{3}k\, {\cal E}_{\alpha}^\ast (k) \Bigg[f_{\alpha}(\mu_{\alpha}^{\ast},T)+\bar f_{\alpha}(\mu_{\alpha}^{\ast},T)\Bigg] +n\,g_\omega\,\omega +
		m_{\sigma}^2{\sigma}^2\Bigg(\frac{1}{2}+\frac{\kappa_{3}}{3!}
		\frac{g_\sigma\sigma}{M} + \frac{\kappa_4}{4!}\frac{g_\sigma^2\sigma^2}{M^2}\Bigg)
		\nonumber\\
		&&
		-\frac{1}{2}m_{\omega}^2\,\omega^2\Bigg(1+\eta_{1}\frac{g_\sigma\sigma}{M}+\frac{\eta_{2}}{2}\frac{g_\sigma^2\sigma^2}{M^2}\Bigg)-\frac{1}{4!}\zeta_{0}\,{g_{\omega}^2}\,\omega^4 + \frac{1}{2}n_{3}\,g_\rho\,\rho - \frac{1}{2}\Bigg(1+\frac{\eta_{\rho}g_\sigma\sigma}{M}\Bigg)m_{\rho}^2\,\rho^{2}
		\nonumber\\
		&&
		-\Lambda_{\omega}\, g_\rho^2\, g_\omega^2\, \rho^2\, \omega^2, \\
		and \nonumber \\ \nonumber
		P & = & \sum_{\alpha=p,n} \frac{2}{3 (2\pi)^{3}}\int d^{3}k\, \frac{k^2}{{\cal E}_{\alpha}^\ast(k)} \Bigg[f_{\alpha}(\mu_{\alpha}^{\ast},T)+\bar f_{\alpha}(\mu_{\alpha}^{\ast},T)\Bigg] - m_{\sigma}^2{\sigma}^2\Bigg(\frac{1}{2} + \frac{\kappa_{3}}{3!}\frac{g_\sigma\sigma}{M} + \frac{\kappa_4}{4!}\frac{g_\sigma^2\sigma^2}{M^2}\Bigg) 
		\nonumber\\
		& &
		+\frac{1}{2}m_{\omega}^2\,\omega^2\Bigg(1+\eta_{1}\frac{g_\sigma\sigma}{M}+\frac{\eta_{2}}{2}\frac{g_\sigma^2\sigma^2}{M^2}\Bigg) + \frac{1}{4!}\zeta_{0}\,{g_{\omega}^2}\,\omega^4 + \frac{1}{2}\Bigg(1+\frac{\eta_{\rho}g_\sigma\sigma}{M}\Bigg)m_{\rho}^2\,\rho^{2}
		\nonumber\\
		& &
		+\Lambda_{\omega}\, g_\rho^2\, g_\omega^2\, \rho^2\, \omega^2.
		\end{eqnarray}
	\end{widetext}
	We will now discuss the liquid-gas phase transition and the computation for critical temperature for such a growth. The self-interaction term of the isovector mesons is mainly responsible for variation in the critical temperature of nuclear matter \cite{Qian_2004}. Two conditions control the essential temperature for the case of SNM, which will be given by \cite{Wang_2000}, 
	\begin{eqnarray} \label{crte}
	\frac{\partial P}{\partial n}\Bigg |_{T=T_C} = \frac{\partial^2P}{\partial n^2}\Bigg |_{T=T_C} = 0.
	\end{eqnarray}
	In other words, we can say that the inflection point of the pressure will determine the critical temperature of the system. Another fundamental quantity of the nuclear matter that influences the EoS and gives adequate information about the gravitational attraction of astrophysical objects is free symmetry energy \cite{GOUDARZI_2018}. Free symmetry energy for the nuclear matter at finite temperature is defined as the difference of free energy per nucleon of pure neutron matter and SNM \cite{Tan_2016}, i.e.,   
	\begin{eqnarray}
	F_{sym}(n,T) = \frac{F(n,T,t=1)}{n} - \frac{F(n,T,t=0)}{n},
	\end{eqnarray}
	where, $F$ is the free energy density given by $F=E-TS$, with $S$, being the entropy density and the expression for the entropy density of the nuclear matter can be written as \cite{Fetter_1971},
	\begin{eqnarray}
	S &=& - \sum_{\alpha=p,n} \frac{2}{(2\pi)^{3}}\int d^{3}k\, \Bigg[f_{\alpha}(\mu_{\alpha}^{\ast},T) \ln{f_{\alpha}(\mu_{\alpha}^{\ast},T)} + \nonumber\\
	& &
	(1-f_{\alpha}(\mu_{\alpha}^{\ast},T)) \ln{(1-f_{\alpha}(\mu_{\alpha}^{\ast},T))}
	+ \bar f_{\alpha}(\mu_{\alpha}^{\ast},T) 
	\nonumber\\
	& &
	\ln{\bar f_{\alpha}(\mu_{\alpha}^{\ast},T)} +(1-\bar f_{\alpha}(\mu_{\alpha}^{\ast},T)) \ln{(1-\bar f_{\alpha}(\mu_{\alpha}^{\ast},T))} \Bigg], \nonumber \\
	\end{eqnarray}
	To obtain the other derivatives of the symmetry energy such as $L_{sym}$, $K_{sym}$ and $Q_{sym}$, we can expand the free symmetry energy ($F_{sym}$) around a variable $\chi$, which in terms of saturation density ($n_{0}$) can be written as $\chi = ({n-n_0})/{3n_0}$. This expansion will take the following form,
	\begin{eqnarray}
	F_{sym}(n,T) &=& F_{sym}(n_0) + L_{sym}\,\chi + \frac{K_{sym}}{2!}\,\chi^2  \nonumber \\  &&
	+ \frac{Q_{sym}}{3!}\,\chi^3 + O(\chi^4).
	\end{eqnarray}
	This expansion will bestow us with the following expression of $L_{sym}$, $K_{sym}$ and $Q_{sym}$ \cite{Chen_2009},
	\begin{eqnarray}
	L_{sym}(n,T) &=& 3n\,\frac{\partial F_{sym}(n,T)}{\partial n}\Bigg |_{n=n_0}, \\ 
	K_{sym} (n,T) &=& 9n^2\,\frac{\partial^2 F_{sym}(n,T)}{\partial n^2}\Bigg |_{n=n_0},\\
	Q_{sym} (n,T) &=& 27n^3\,\frac{\partial^3 F_{sym}(n,T)}{\partial n^3}\Bigg |_{n=n_0},
	\end{eqnarray}
	with $n_{0}$ being the saturation density. Thermal index ($\Gamma_{th}$), which is a crucial quantity to explore the dynamics of core-collapse supernovae explosion \cite{Yasin_2020}, can be extracted from the finite temperature EoS with a simple expression \cite{Carbone_2019},
	\begin{eqnarray}
	\Gamma_{th} &=& 1 + \frac{P_{th}}{E_{th}},
	\end{eqnarray}
	where $E_{th}$ and $P_{th}$ will be given by $E_{th} = E(T) - E(0)$ and $P_{th} = P(T) - P(0)$, with $E(T)$ and $P(T)$ being the energy density and pressure at temperature $T$.
	\subsection{Neutrino Emissivity}\label{Cool} 
	This section will discuss the theoretical formalism for neutrino emissivity ($Q$), which is usually triggered by the direct Urca process and requires a threshold fraction of proton in the core of neutron star to operate \cite{LATTIMER_2007}. Lattimer {\it et al.}, \cite{Lattimer_1991} calculated a simple formula for neutrino emissivity for $\beta$-equilibrium nuclear matter using the non-relativistic formalism. Here, we have used the RMF formalism to calculate the expression for neutrino emissivity, which made some improvements later, and modify the formula by taking into account the proton recoil and the difference in the neutron-proton potential energies \cite{LEINSON_2001}. We presume a dense matter system of protons and neutrons along with the electrons to understand the effects of direct Urca process $n \longrightarrow p + e^- + \bar\nu_e$ and $p + e^- \longrightarrow n + \nu_e$, on the neutrino emissivity and the cooling mechanism of the newly born star. We redefine our Lagrangian accordingly, which now will be given by,
	\begin{eqnarray}
	{\cal L}_{total} &=& {\cal L} + \bar\phi\,(i\gamma_{\mu} \partial^{\mu} - m_e)\phi,
	\end{eqnarray}
	here, $\phi$ is the wave-function and $m_{e}$ is the mass of the electron, and ${\cal L}$ corresponds for the Lagrangian stated in sec. \ref{RMF}. Energy density and pressure will be modified accordingly for the assumptive system and will be given by,
	\begin{eqnarray}\label{edt}
	E_{total} &=&  E + \frac{2}{(2\pi)^{3}}\int d^{3}k\, {\cal E}_{e} (k)\, [f_{e}(T)+\bar f_{e}(T) ], \\
	P_{total} &=&  P + \frac{2}{3(2\pi)^{3}}\int d^{3}k\, \frac{k^2}{{\cal E}_{e}}\, \Big[f_{e}(T)+\bar f_{e}(T)\Big]. 
	\end{eqnarray}
	Here $k_{e}$ is the electron's momentum, $f_{e}(T)$ is the thermal distribution function for electron at temperature $T$ and ${\cal E}=\sqrt{k^2_e + m^2_e}$ is the energy of the electron. E and P are the energy density and pressure derived in sec. \ref{SNM} for the Lagrangian $\cal L$. To calculate neutrino emissivity expression using relativistic formalism, we follow the procedure suggested by L. B. Leinson and A. Perez. The detailed calculation for the formula of neutrino emissivity (Q) in the relativistic framework applying mean-field approximation, which is used here, can be found in the reference \cite{LEINSON_2001,LEINSON_2002}. This derived expression is valid only for low temperature expansion. We have also used the calculated expression for neutrino emissivity $Q$ in our previous work for NL3, G3 and IOPB-I equation of states \cite{Kumar_2020}. The final expression for Q is,
	\begin{widetext}
		\begin{eqnarray}
		Q &=& \frac{457\pi}{10080} G_F^{2}\,C^2\,T^6\,\Theta(k_e + k_p - k_n) \Bigg\{\left(C_A^2-C_V^2\right) M_p^\ast\, M_n^\ast\, {\cal E}_e + \frac{1}{2} \left(C_V^2+C_A^2\right) \bigg[4\,{\cal E}_n\, {\cal E}_p\,  {\cal E}_e - \left({\cal E}_n - {\cal E}_p\right) 
		\nonumber  \\[3mm]  & &  \null 
		\left(\left({\cal E}_n + {\cal E}_p \right)^2 - k_e^2\right) \bigg] + C_V\, C_M\, \frac{\sqrt{M_p^\ast\,M_n^\ast}} {M} \left[2 \left({\cal E}_n - {\cal E}_p \right) k_e^2 - \left(3 \left(  {\cal E}_n - {\cal E}_p \right)^2 - k_e^2 \right) {\cal E}_e \right] + C_A 
		\nonumber  \\[3mm]  & &  \null  
		\left(C_V + 2\frac{\sqrt{M_p^\ast M_n^\ast}}{M} C_M \right) \left({\cal E}_n + {\cal E}_p \right) \left(  k_e^2 - \left({\cal E}_n + {\cal E}_p \right)^2\right) + C_M^2 \frac{1}{4M^2} \bigg[8M^{\ast 2} \left( {\cal E}_n - {\cal E}_p \right) \left(k_e^2 - \left( {\cal E}_n - {\cal E}_p\right) {\cal E}_e \right)
		\nonumber  \\[3mm]  & &  \null
		+ \left(k_e^2 - \left( {\cal E}_n - {\cal E}_p\right)^2 \right) \left(2{\cal E}_n^2 + 2{\cal E}_p^2 - k_e^2 \right) {\cal E}_e - \left(k_e^2 - \left( {\cal E}_n - {\cal E}_p\right)^2 \right) \left({\cal E}_n + {\cal E}_p \right)\left(2{\cal E}_n - 2{\cal E}_p - {\cal E}_e \right)  \bigg] \Bigg\},  
		\end{eqnarray}
	\end{widetext}
	where $G_F = 1.166\times10^{-11}$ MeV$^{-2}$ is defined as the Fermi coupling constant, $C$ = 0.973 as the Cabibbo factor, $C_V$ = 1 and $C_A$ = 1.26 represent the vector and axial-vector constants respectively and the constant $C_M$ = 3.7 stands for the weak magnetism effects. The condition necessary for Urca processes to operate will be given by $\Theta(k_e + k_p - k_n)$ = 1, if $k_e + k_p - k_n \geq 0$ and zero otherwise, where $k_e$, $k_p$ and $k_n$ are the Fermi momenta of electron, proton and neutron respectively.
	\subsection{Proto-Neutron Star}\label{PNS}
	\noindent
	The plethora of information on the dense nuclear matter at finite temperature endows us with enough stimulation to investigate the newly born dense star properties in detail. In this work, we mainly focus on modifying the M-R profile of the PNS with the temperature at constant entropy. The M-R profile for a proto-neutron star can be obtained by solving the Tolman-Oppenheimer-Volkoff (TOV) equations \cite{Tolman_1939} with EoS as input. Many factors affect the profile and internal properties of a proto-neutron star during its evolutionary stage. One of the elements is neutrino emission that we discussed in the previous section. As we know that the newly born dense star cools down by emitting a large number of neutrinos, however, some of the neutrinos get trapped inside the core of the neutron star due to its small mean free path. To perceive the effect of trapped neutrino on the PNS structure, we modified our Lagrangian and EoS accordingly. The Lagrangian to study the M-R profile of the newly born star at finite temperature will be given by
	\begin{eqnarray}
	{\cal L}_{PNS} &=& {\cal L} + \sum_{l=e,\mu}\bar\phi_{l}\,(i\gamma_{\alpha} \partial^{\alpha} - m_{l})\phi_{l} + \bar\phi_{\nu_e}\,(i\gamma_{\alpha} \partial^{\alpha})\phi_{\nu_e}, \nonumber \\
	\label{LPNS}
	\end{eqnarray}
	where, $m_{l}$ and $\phi_{l}$ stands for the mass and wave-function of leptons (i.e. electrons and muons). The last term in Eq. (\ref{LPNS}) is responsible for the neutrino trapping, and $\phi_{\nu_e}$ corresponds for the wave-function of neutrino. The inevitable conditions required for the stability of a PNS are $\beta$-equilibrium and charge neutrality. The neutrons inside the neutron star will eventually follow the $\beta$ decay process to maintain the equilibrium and to maintain the charge neutrality muons will appear when the chemical potential of electron reaches the value of muon rest mass i.e. $106 MeV$. The charge neutrality and the $\beta-$ equilibrium conditions for a given nucleon density ($n$) will be given by,
	\begin{eqnarray} \label{eqm}
	n_{p} &=& n_{e} + n_{\mu}, \nonumber \\
	\mathrm{and} \nonumber \\
	\mu_{n} &=& \mu_{p} + (\mu_{e} - \mu_{\nu_e}), \nonumber \\
	\mu_{e} &=& \mu_{\mu},
	\end{eqnarray}
	where $\mu_{n}, \mu_{p}, \mu_{e}, \mu_{\mu}$ and $\mu_{\nu_e}$ describe the chemical potentials of the neutron, proton, electron, muon and neutrino respectively; $n_{p}$, $n_{e}$ and $n_{\mu}$ are the number densities of proton, electron and muon. The particle fractions of neutrons and protons will depend on both charge neutrality and the $\beta-$ equilibrium condition. The self consistent numerical solution of Eq.(\ref{eqm}) will set the fraction of neutron, proton, electron and muon number density for a given baryon density in a neutron star. The energy density and pressure terms for lepton terms at finite temperature will given by,
	\begin{eqnarray} \label{elpl}
	{\epsilon}_{l} &=& \sum_{l=e,\mu} \frac{1}{4\pi^{3}} \int \sqrt{k^{2} + m_{l}^{2}}\, [f_{l}(T)+\bar f_{l}(T)] d^{3}k,  \\
	{P}_{l} &=& \sum_{l=e,\mu} \frac{1}{12\pi^{3}} \int \frac{k^{2} [f_{l}(T)+\bar f_{l}(T)]d^{3}k}{\sqrt{k^{2} + m_{l}^{2}}},
	\end{eqnarray}
	where $f_{l}(T)$ and $\bar f_{l}(T)$ are the distribution functions of leptons and anti-leptons respectively. Now, the energy density and pressure for the $\beta-$ equilibrated PNS along with the effect of trapped neutrinos can be derived by following the same procedure discussed in sec. \ref{SNM} and references \cite{Zhou_2017, Pons_1999}, and stated as,
	\begin{eqnarray} \label{EoSS}
	E_{PNS} &=&  E + {\epsilon}_{l} + \sum_{\nu_e} \Bigg(\frac{7\pi^2}{120} + \frac{\mu^2_{\nu_e}T^2}{12} - \frac{\mu^4_{\nu_e}}{24 \pi^2} \Bigg), \\
	P_{PNS} &=&  P + P_{l} \nonumber \\
	& + &\sum_{\nu_e} \frac{1}{360} \Bigg( 7\pi^2 T^{4} + 30 \mu^2_{\nu_e} T^2 + \frac{15 \mu^4_{\nu_e}}{\pi^2} \Bigg),
	\end{eqnarray}
	where $E$ and $P$ are the energy density and pressure defined in sec. \ref{SNM}, $\mu_{\nu_e}$ stands for the chemical potential of neutrino, and the last terms are the consequences of neutrino trapping at temperature $T$. 
	We can calculate the mass and radius for a PNS from two different approaches, i.e., either by keeping the temperature constant throughout the or by fixing the nucleons' entropy inside the star. In our previous work \cite{Kumar_2020}, we observe that the constant entropy approach is more suitable and appropriate to explore the effect of temperature on the M-R profile of neutron stars. In this work, we analyze the effect of temperature on the proto-neutron star's mass and radius by exploring the constant entropy EoS. The expression for the entropy per nucleon of the proto-neutron star can be stated as \cite{Pons_1999,PRAKASH_1997},
	\begin{eqnarray}
	S &=& \frac{E_{PNS} + P_{PNS} - \sum_{\alpha=p,n} n_{\alpha}. \mu_{\alpha}}{nT}.
	\end{eqnarray}
	To proceed with further calculations, we need to study how diverse is the lepton fraction in the early evolutionary stages of the newly born proto-neutron star. The lepton fraction for the proto-neutron star can be defined as $Y_{L}=\frac{n_{e} + n_{\nu_e}}{n}$ and we fix its value to $0.4$ for our further perusal \cite{Zhou_2017, PRAKASH_1997}. Now, the M-R profile of the star can be easily obtained by solving the TOV equation. The TOV equations for the static isotropic proto-neutron star can be written as,
	\begin{eqnarray}
	\frac{d P_{PNS}(r)}{d r} &=&  -\frac{[E_{PNS}(r)+P_{PNS}(r)]}{r^2\Big(1-\frac{2M(r)}{ r}\Big)} \nonumber \\
	&&
	\times [M(r)+{4\pi r^3 P_{PNS}(r)}], \\
	\frac{d M(r)}{d r} &=& 4\pi r^2 {E_{PNS}(r)}.
	\end{eqnarray}
	Here, $M(r)$ is defined as the mass of the neutron star at radius r and the boundary conditions to solve these equations are $P_{PNS}(R) = 0$, for a particular choice of central density $\rho_c = \rho(0)$. The neutron star inner and outer crust EoS can also be added along with the core energy and pressure derived from Eqs. (31) and (32) \cite{1971ApJ...170..299B}.
	
	\section{Results and Discussion}\label{result}
\begin{figure}[h]
		\centering
		\includegraphics[width=0.47\textwidth]{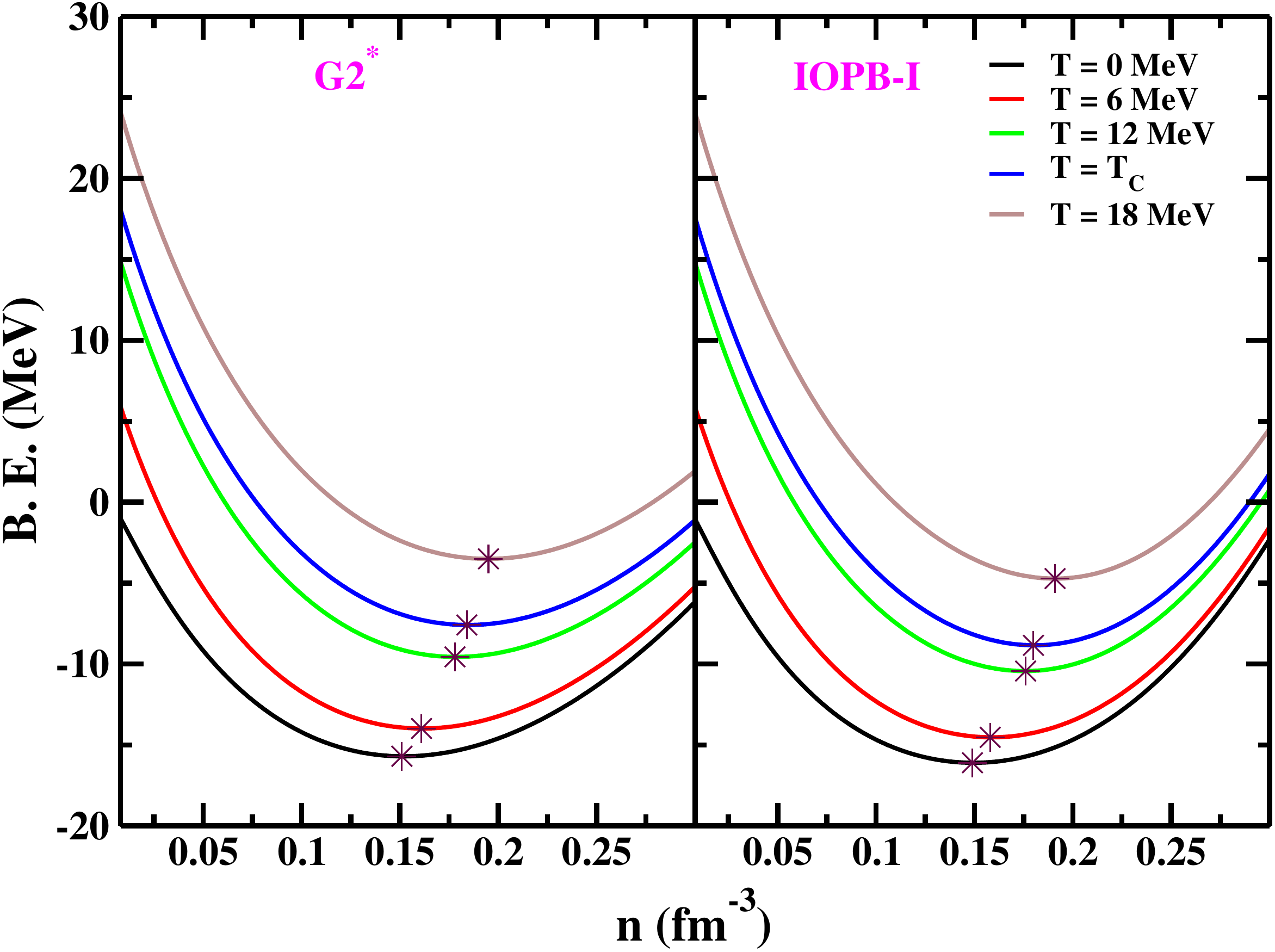}
		\caption{(colour online) Binding Energy as a function of nucleon density for SNM ($t=0$) at different temperature. The left panel shows the results for G2$^{*}$ and right panel for IOPB-I parameter set. The star symbol indicates the minima of the curve and the corresponding saturation density.}
		\label{BE}
	\end{figure}

\begin{figure}[h]
		\centering
		\includegraphics[width=0.47\textwidth]{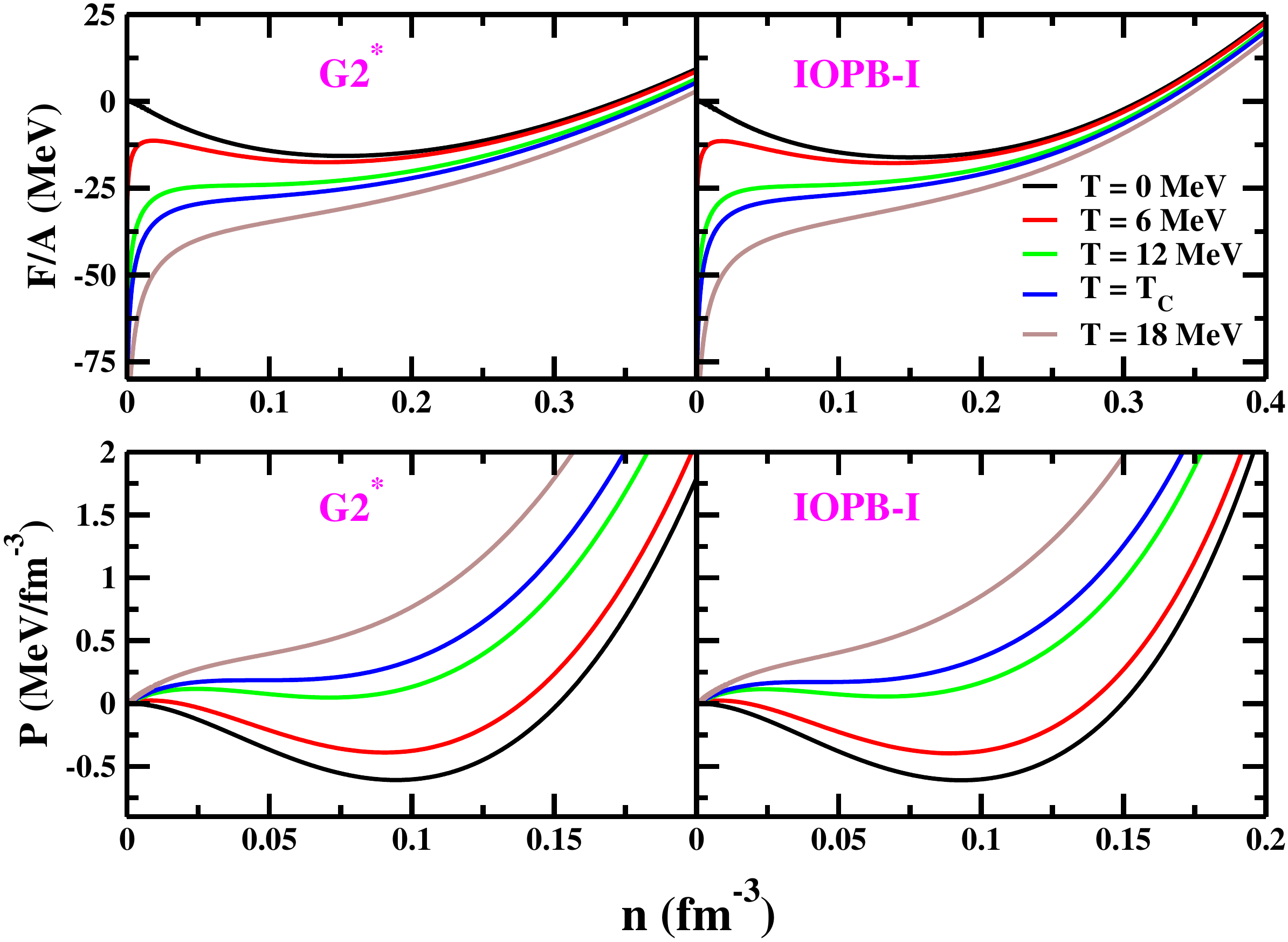}
		\caption{(colour online) Free energy and pressure as a function of nucleon density for SNM ($t=0$) at different temperature. The left panel shows the results for G2$^{*}$ and right panel for IOPB-I parameter set.}
		\label{FP}
	\end{figure}

	\noindent
	In the last few decades, RMF formalism emerges as one of the most important and prominent theories capable of interpreting the finite nuclei results of heavy-ion collision experiments and the data obtained from the astrophysical observations adequately. A lot of RMF parameter sets has been developed in the last few years, which endue us with different types of equations of states, like, NL3 \cite{Lalazissis_1997}, the most familiar and fundamental RMF parameter set, provides the stiffest EoS and others like FSU-Gold \cite{Todd_2005}, IU-FSU \cite{Carbone_2019}, G3 \cite{Kumar_2020} etc. dominates the softer region of EoS. Later various theoretical studies put some constraints on the RMF parameter sets' consistency using the experimental and observational data. Some of the RMF parameter sets failed to elucidate the experimental studies and have been considered incompatible for a more consistent study of astrophysical objects. In this work, we used G2$^{*}$ and IOPB-I parameter forces to explore the thermal properties of the nuclear matter and the neutron star. However, G2$^{*}$ is a well-informed and consistent parameter set that satisfy all the constraints set by the observational studies \cite{Lourenco_2019}, and IOPB-I is the recently developed parameter force by our group, which has also been recognized as a compatible RMF set in the theoretical studies \cite{Bharat_2018}. The numerical values for all the coupling constants of G2$^{*}$ and IOPB-I parameter sets are provided in the upper portion of Table \ref{table1}. The saturation nuclear matter properties at $T=0$ have also been presented along with the available experimental data in the lower segment of the table. \\
	The variation of the binding energy (B.E.) of SNM with temperature is shown in Fig. \ref{BE}. We observe that with the increase in temperature, the saturation density of the SNM also increases. The matter gets saturated at a more significant density for higher temperatures. Also, we noticed that the system becomes less stable at the higher temperature, as there is an increase in the B.E. of the system with temperature. The presumed parameter sets are consistent with this observation of higher saturation energy and lower binding energy of the course with a temperature rise. The free energy per
	particle F/A at different temperatures is shown in Fig. \ref{FP} and are in good agreement
	with the microscopic Brueckner-Hartree-Fock (BHF) calculated results in the ref. \cite{PhysRevC.93.035806}. Another important information that we can avail from the EoS of warm SNM is the determination of critical temperature ($T_{C}$) for a liquid-gas phase transition. The onset of liquid-gas phase transition starts with the flattening of the pressure curve, which is plotted as a function of nucleon density at different temperatures in the lower panel of Fig. \ref{FP}. We found the value of critical temperature for both the parameter sets from the inflexion point of the pressure curve by applying the conditions as discussed in Eq. (\ref{crte}). We obtained the value of $T_{C}$ as $14.10$ and $13.82$ MeV for G2$^{*}$ and IOPB-I parameter sets, respectively.     
	
\begin{figure}[h]
		\centering
		\includegraphics[width=0.47\textwidth]{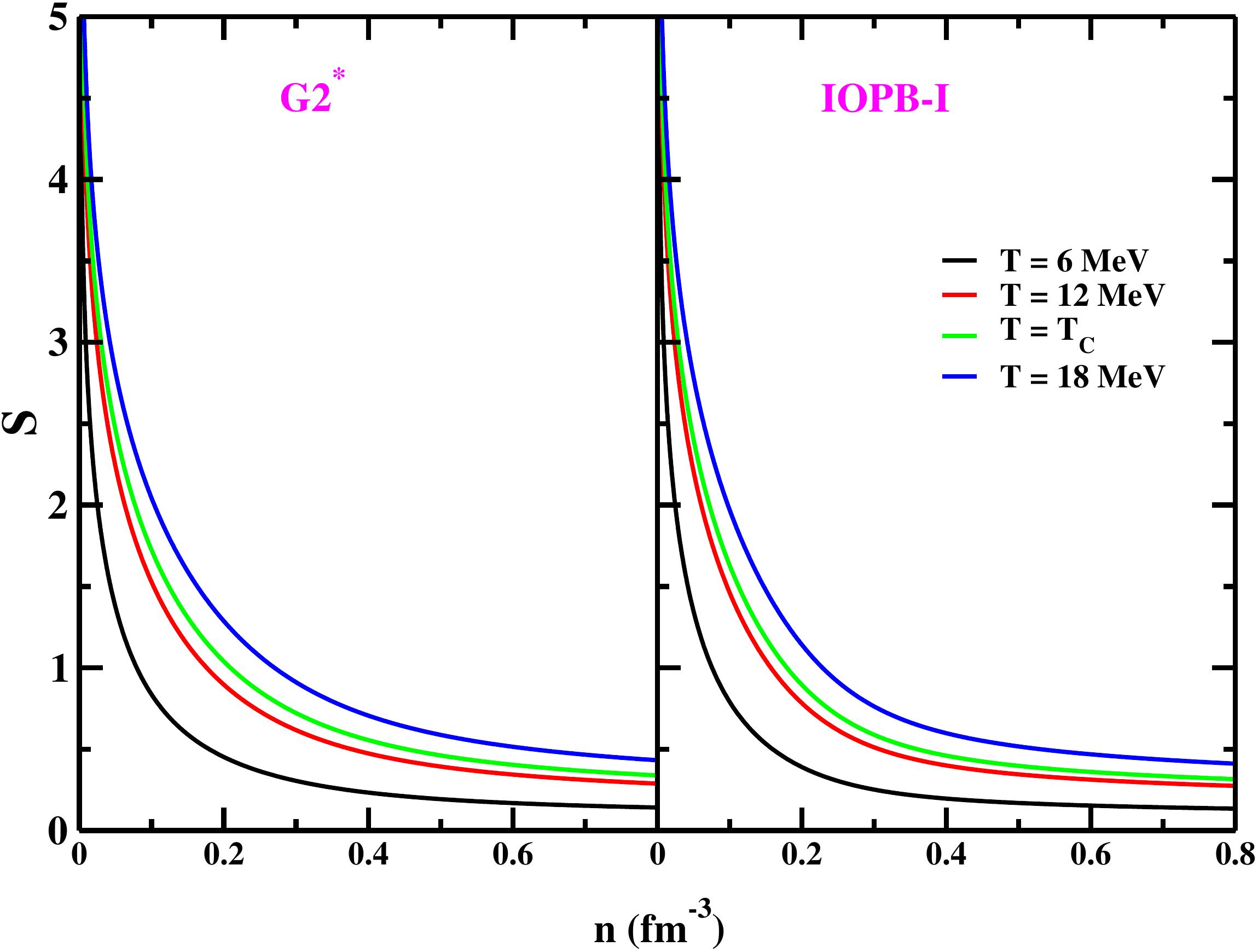}
		\caption{(colour online) Entropy ($S$) of SNM ($t=0$) as a function of nucleon density at different temperatures. The left panel shows the results for G2$^{*}$ and the right panel for IOPB-I parameter set.}
		\label{Ent}
	\end{figure}

\begin{figure}[h]
		\centering
		\includegraphics[width=0.47\textwidth]{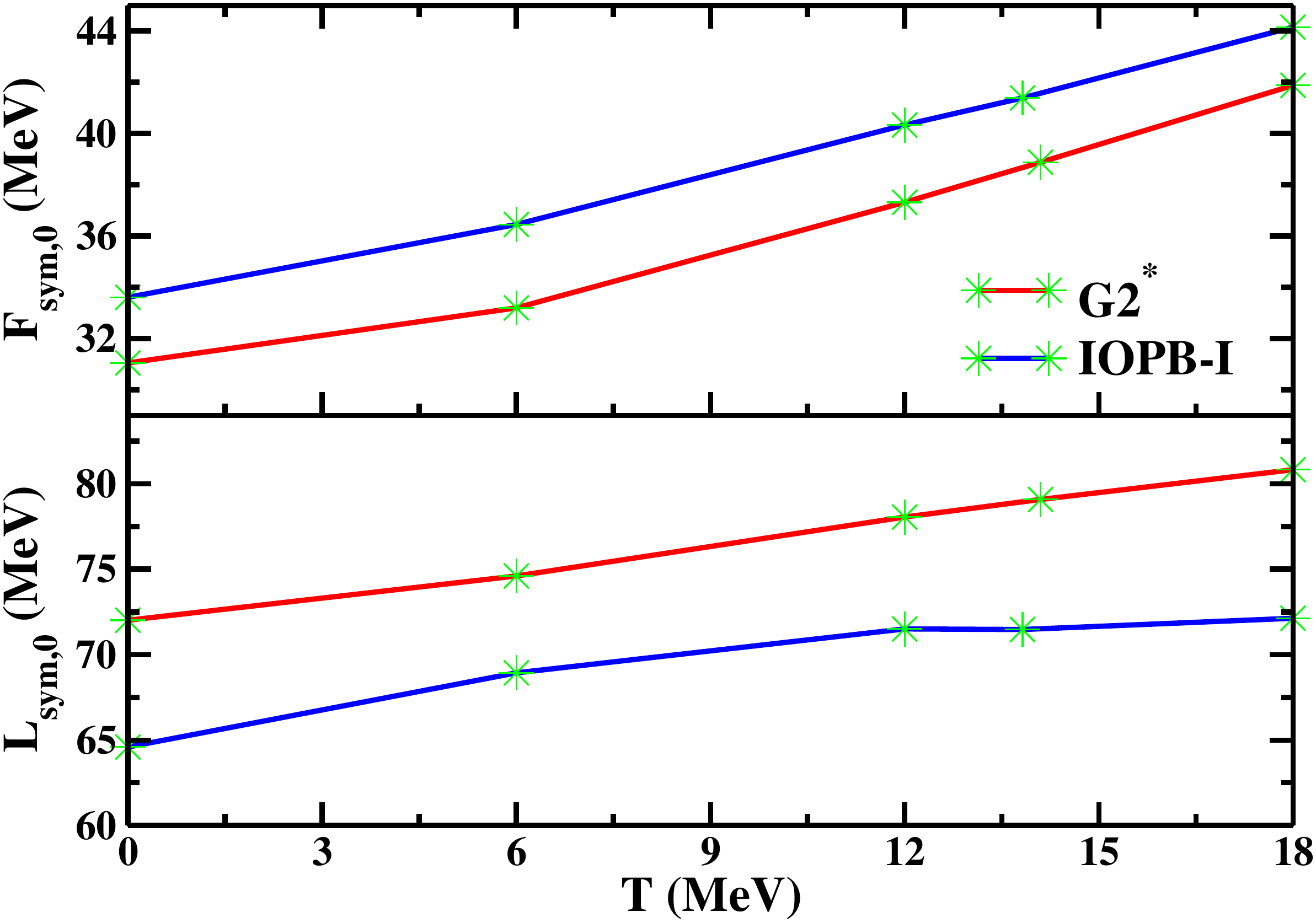}
		\caption{(colour online) Variation of symmetry energy and slope parameter at saturation density ($F_{sym,0}$ and $L_{sym,0}$) with temperature for SNM.}
		\label{Lvari}
	\end{figure}

\begin{figure}[h]
		\centering
		\includegraphics[width=0.47\textwidth]{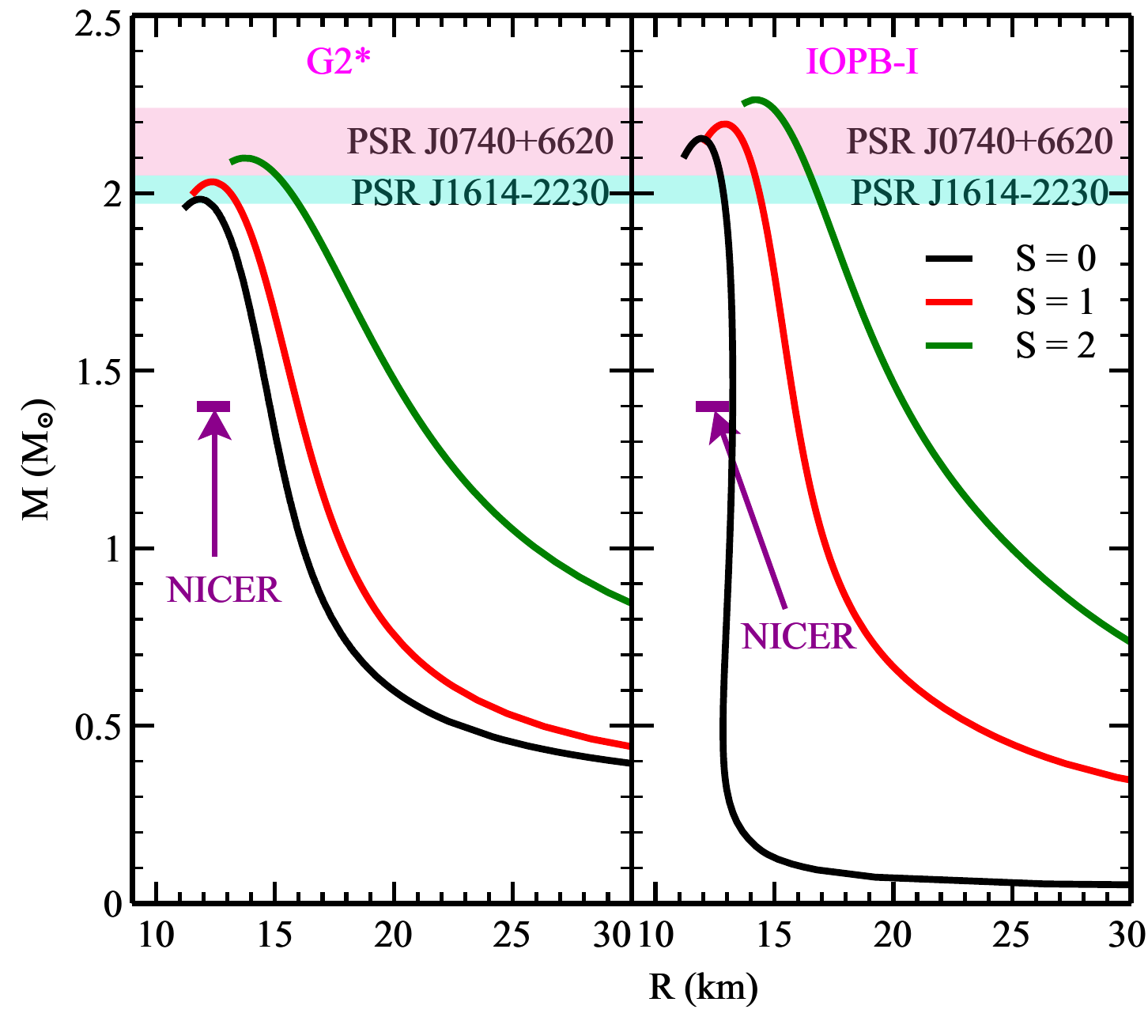}
		\caption{(colour online) M-R profile of proto-neutron star at constant entropy for G2$^{*}$ (left panel) and IOPB-I (right panel) parameter sets. }
		\label{M-R}
	\end{figure}

	\begin{table}
		\caption{The coupling constants and the nuclear properties for G2$^{*}$ \cite{Sulaksono_2005, PhysRevC.90.055203} and IOPB-I \cite{Bharat_2018} parameter sets at saturation. The coupling constants have no units, except $k_3$ and $n_{0}$ which are in fm$^{-1}$ and fm$^{-3}$ respectively. The parameters $n_{0}$, $B.E.$, $F_{sym,0}$ and $L_{sym,0}$ are given at saturation for T = 0 K and in MeV units in the lower panel. The references are $[a]$,$[b]$, $[c]$ $\&$ $[d]$ \cite{Zyla_2020}, $[e] $\&$ [f]$ \cite{Bethe_1971}, $[g] $\&$ [h]$ \cite{DANIELEWICZ_2014}.}
		\scalebox{1.2}{
			\begin{tabular}{cccccccccc}
				\hline
				\hline
				\multicolumn{1}{c}{Parameter}
				&\multicolumn{1}{c}{G2$^{*}$}
				&\multicolumn{1}{c}{IOPB-I}
				&\multicolumn{1}{c}{Empirical/Expt. Value}\\
				\hline
				$m_{\sigma}/M$ & 0.554 & 0.533 &0.426 -- 0.745 $[a]$\\
				$m_{\omega}/M$  &  0.833  &  0.833 & 0.833 -- 0.834 $[b]$  \\
				$m_{\rho}/M$  &  0.812  &  0.812 & 0.825 -- 0.826 $[c]$\\
				$g_{\sigma}/4 \pi$  &  0.835  &  0.827  & \\
				$g_{\omega}/4 \pi$  &  1.016  &  1.062  &\\
				$g_{\rho}/4 \pi$  &  0.938 &  0.885   & \\
				$k_{3} $   &  3.247 &  1.496  &  \\
				$k_{4}$  &  0.632  & -2.932   &\\
				$\zeta_{0}$  &  2.642 &  3.103  &   \\
				$\eta_{1}$  &  0.650 &  0.0  & \\
				$\eta_{2}$  &  0.110 &  0.0  &  \\
				$\eta_{\rho}$  &  4.490 &  0.0 & \\
				$\Lambda_{\omega}$  &  0.0 &  0.024  &  \\
				\hline
				$n_{0}$ & 0.154 & 0.149 & 0.148 -- 0.185 $[e]$\\
				$B.E.$ & -16.07 & -16.10 & -15.00 -- 17.00 $[f]$\\
				$F_{sym,0}$ & 30.39 & 33.35 & 30.20 -- 33.70 $[g]$\\
				$L_{sym,0}$ & 69.68 & 61.76 & 35.00 -- 70.00 $[h]$ \\
				\hline
				\hline
			\end{tabular}
		}
		\label{table1}
	\end{table}
	\begin{table}
		\caption{The critical values of temperature, pressure, and density for symmetric nuclear matter using G2$^{*}$ and IOPB-I parameter sets. The theoretically calculated values for NL3 and G3 parameter sets and the available experimental data have also been presented.}
		\scalebox{1.2}{
			\begin{tabular}{cccccccccc}
				\hline
				\hline
				\multicolumn{1}{c}{Parameter}
				&\multicolumn{1}{c}{$T_{C}$}
				&\multicolumn{1}{c}{$P_{C}$}
				&\multicolumn{1}{c}{$n_{C}$}\\
				-- & ($MeV$) & $(MeV/fm^{3})$ & $(fm^{-3})$ \\
				\hline
				G2$^{*}$ & 14.10 & 0.185 & 0.046 \\
				IOPB-I & 13.82 & 0.171 & 0.042  \\
				NL3 \cite{Kumar_2020} & 14.60 & 0.191 & 0.053 \\
				G3 \cite{Kumar_2020} & 15.37 & 0.162 & 0.062 \\
				Exp.1  \cite{Elliott_2013}  & 17.90$\pm$0.4 & 0.31$\pm$0.07 & 0.06$\pm$0.01 \\
				Exp.2 \cite{Li_1994} & 13.10$\pm$0.6  & -- & 0.05$\pm$0.01\\
				\hline
				\hline
			\end{tabular}
		}
		\label{table2}
	\end{table}
	The corresponding critical density ($n_{C}$) and pressure ($P_{C}$) can also be estimated with the help of the EoS at $T_{C}$ \cite{Yang_2019}. The critical values for G2$^{*}$ and IOPB-I parameter sets along with the data from various theoretical and experimental studies is given in table \ref{table2}.
	We observe that the theoretically calculated values for critical temperature are slightly lower than that of the Exp.1 \cite{Elliott_2013}. However, the required temperature and density of Exp.2 \cite{Li_1994} are in good agreement with the predicted results. We realize that the mismatch in critical pressure is impeccable precision at much lower density, requiring a detailed study of low, dense matter. Fig. \ref{Ent} depicts the variation of entropy for both the adopted parameter sets. The entropy of the symmetric nuclear matter decreases exponentially with the increase in density. At lower density, the nucleons in the system are loosely packed and have more randomness. As the density increase, the system is tightly bound which considerably decrease the entropy of the particles. Also, as expected, entropy is more significant for high temperatures.
	\begin{figure}[h]
		\centering
		\includegraphics[width=0.5\textwidth]{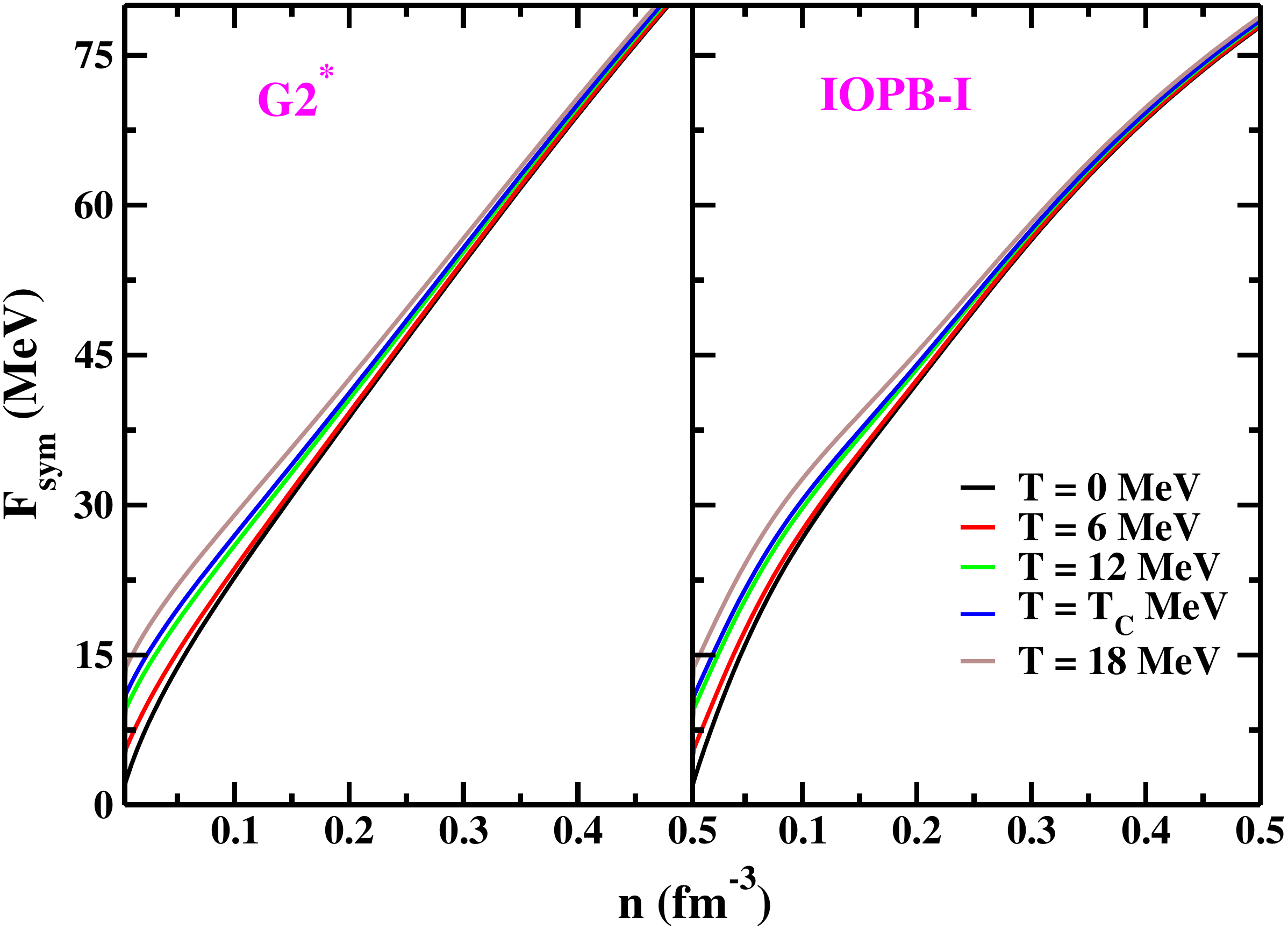}
		\caption{(colour online) Free Symmetry Energy ($F_{sym}$) of SNM ($t=0$) as a function of nucleon density for a definite range of temperature. The left and right panels show the results for G2$^{*}$ and IOPB-I parameter sets respectively.}
		\label{symm}
	\end{figure}
	The change in the free symmetric energy as a function of density with temperature for both the assumptive parameter sets is depicted in Fig. \ref{symm}. In recent years, the density dependence of symmetry energy was considered an essential ingredient to explore the structure of finite nuclei and explain the constraints set by heavy-ion collision experiments and astrophysical observational data. From the linear increment in the free symmetry energy with density for the whole assumed temperature range, we deduce that an abundance of initiation energy is required for neutron-proton conversion in highly dense systems \cite{Li_2011}. We notice quite a significant change in the magnitude of $F_{sym,0}$ and $L_{sym,0}$ with the increase in temperature at saturation density (Fig. \ref{Lvari}).
	However, there is an increase in symmetry energy at saturation density with the rise of temperature. In recent years, there had been many predictions and theoretical calculations regarding the range of symmetry energy. We found that the magnitude of the symmetry energy determined by both the parameter sets befall in the approved range by various studies only for lower region of temperature \cite{Carbone_2019,Burrello_2019,Agrawal_2014}.
	\begin{figure}[h]
		\centering
		\includegraphics[width=0.47\textwidth]{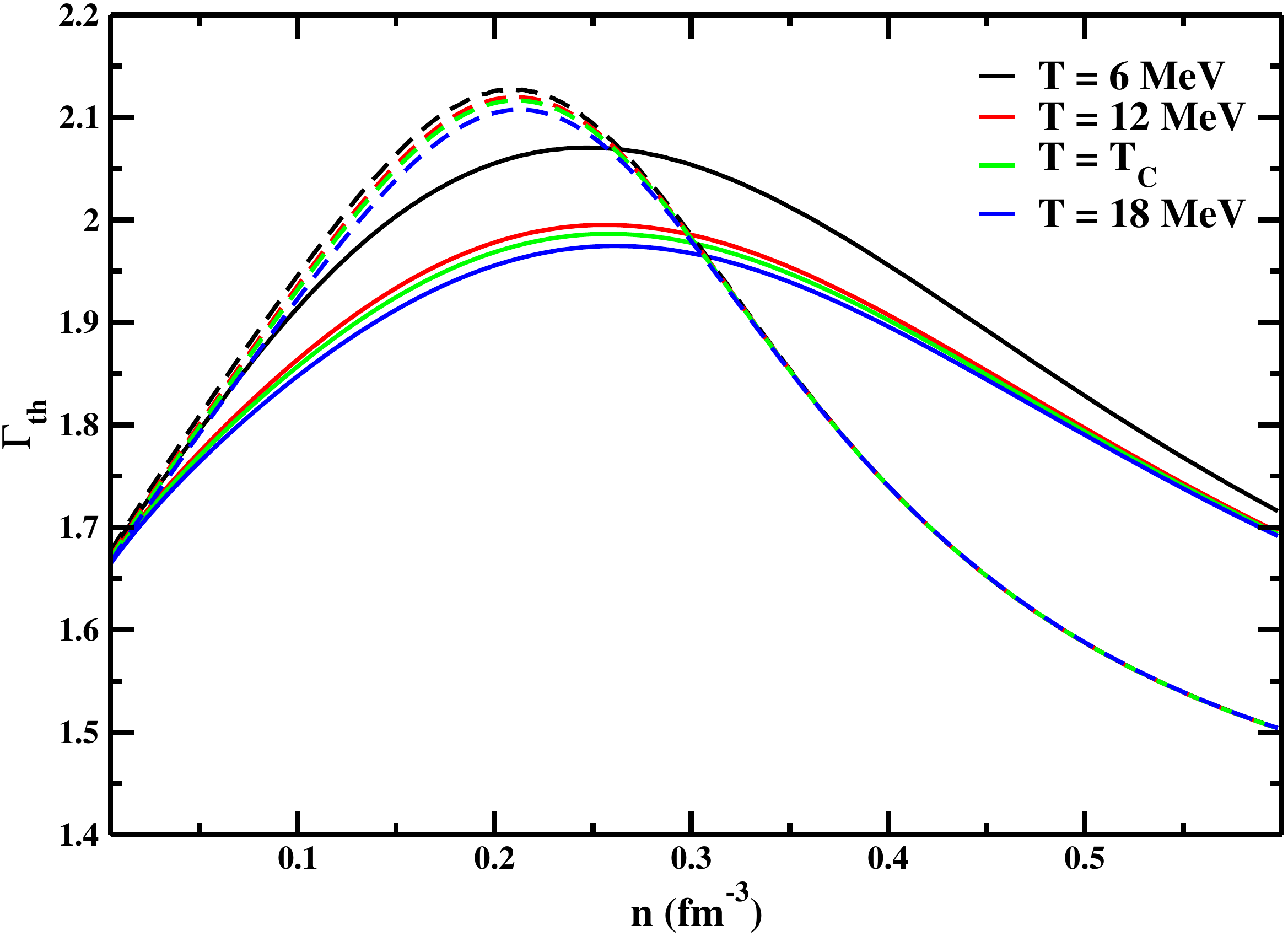}
		\caption{(colour online) Variation of Thermal Index ($\Gamma_{th}$) as a function of nucleon density for G2$^{*}$ (solid) and IOPB-I (dotted) parameter sets.}
		\label{TI}
	\end{figure}
	The spectrum of slope parameter for IOPB-I parameter set supports the empirical data (Table \ref{table1}) more strongly for whole range of temperature, contrary to the G2$^{*}$ outcome, which plays a vital role in determining the neutron skin thickness for finite nuclei \cite{TSANG_2019}. One of the most essential and crucial quantities to recognize the thermal effects on the supranuclear dense matter, which primarily leads the interpretative exposition for the merger dynamics, is the thermal index. Some of the definite facts about the nature of the thermal index can help us constrain the EoS and parameter sets. For instance, the adiabatic thermal index for a non-relativistic ideal gas is a well-known quantity and equal to 5/3, reflected in Fig. \ref{TI} at low densities. We sensate that the matter with low density for the whole temperature range behaves as a non-relativistic ideal gas and approaches the value 5/3 for both the parameter sets. At very high temperatures, the system behaves like a relativistic gas, and in that case, Taub's inequality can be imposed to constrain the value of thermal index  \cite{Carbone_2019}. As the density increase, $\Gamma_{th}$ approaching the value 4/3 agitates the fabrication for a very high thermal pressure to maintain the equilibrium in the dense stars \cite{Bauswein_2010}. The thermal index should approach the value 4/3 for high temperature and density to satisfy the kinetic theory's consistency, which the IOPB-I parameter proves quite satisfactorily. Studies show that the magnitude of the thermal index can constrain the band for effective mass, ultimately controlling the rapidity of contraction and time delay of a massive star \cite{Yasin_2020, Bauswein_2010}.
	\begin{figure}[h]
		\centering
		\includegraphics[width=0.47\textwidth]{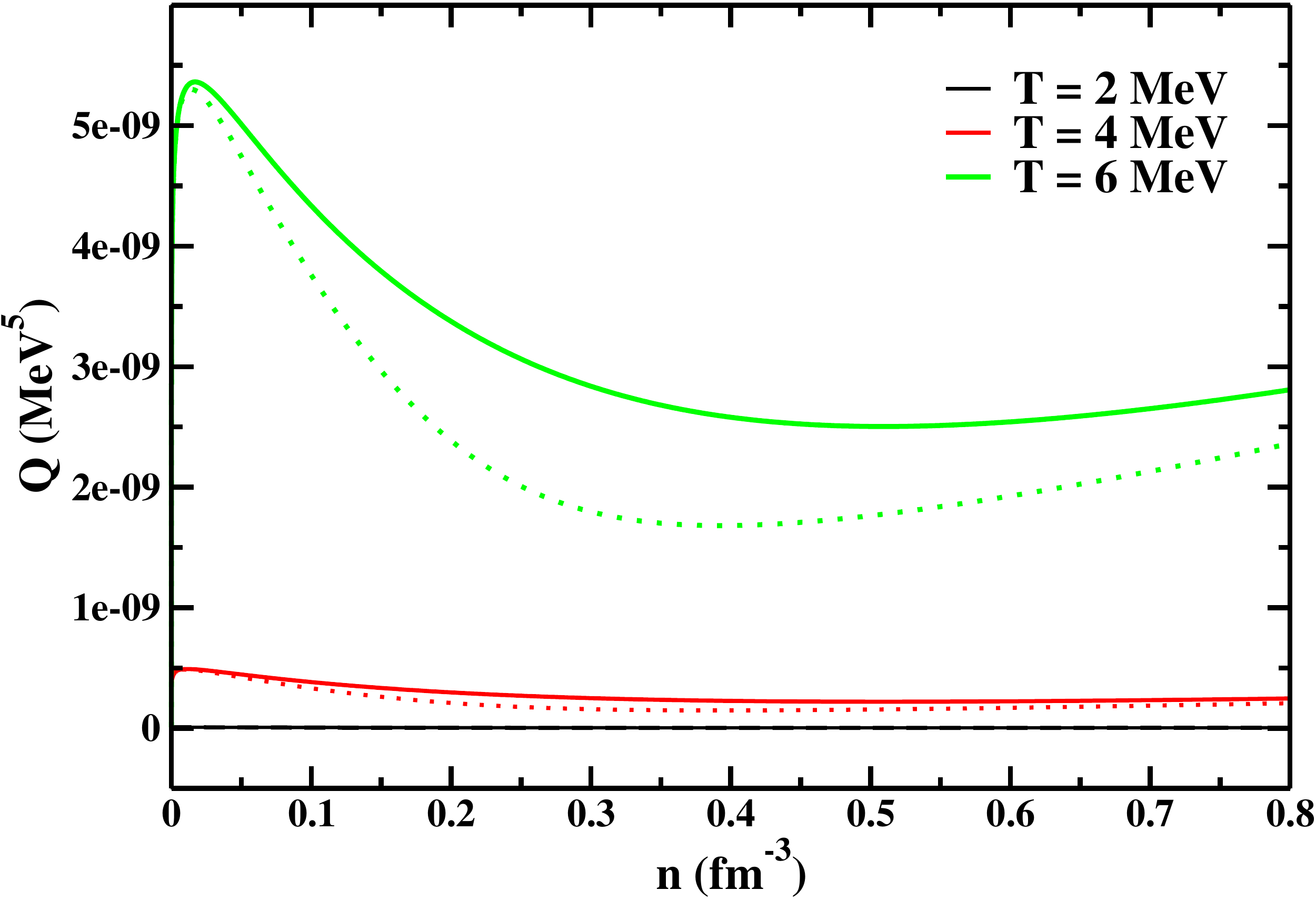}
		\caption{(colour online) Neutrino emissivity at different temperatures as a function of nucleon density for G2$^{*}$ (solid) and IOPB-I (dotted) parameter sets. }
		\label{cool}
	\end{figure}
	The neutrino emissivity of hot dense matter depicted in Fig. \ref{cool} procures a lot of information about the cooling procedure of supernovae remnants. The solid lines in Fig. \ref{cool} represent the calculated results for the G2$^{*}$ parameter set, and the dotted lines stand for the IOPB-I parameter set. From the graph, we can observe that the magnitude of $Q$ decreases quite steeply with a slight temperature change. An enormous difference in the magnitude of the neutrino emissivity magnitude at $T=2$ and $T= 6\,MeV$ for both the parameter sets clearly declare that the cooling rate of the dense matter in its initial stage of evolution is dominated only by the neutrino emission and neutrino emissivity thermally stables the matter within a fraction of seconds \cite{Brown_2018,Yakovlev_2011}. Once the matter cools down to a specific temperature, then the loss of heat takes place mainly through photonic emission, and the trapping of neutrinos effectively suppresses its emission process. Another important conclusion that we draw from this calculation is that generally, lighter remnants of the supernovae explosion cools more rapidly through the neutrino emission than the heavier ones. As you can see from Figs. \ref{cool} and \ref{M-R} that G2$^{*}$ predicts a lower mass for the newly born dense star and higher neutrino emissivity as compare to the IOPB-I parameter set. So, we concluded that the neutrino emission through the direct Urca process in proton-neutron star is enhanced by the increment of mass~\cite{Kumar_2020}.    
	The M-R profile of the newly born proto-neutron star and trapped neutrino effect have also been calculated with a constant entropy approach using TOV equations and is depicted in Fig. \ref{M-R} and also listed in Table \ref{table3}. Some theoretical studies postulated that the maximum mass of a PNS is considerably affected by the system's entropy per baryon in its evolutionary stage. To examine the impact of entropy on the M-R profile, we calculated the mass and radius of PNS for three different fixed entropy per nucleon, i.e., S = 0, 1, and 2. We notice that with an increase in density, the pressure of the PNS also increases, which results in the increment of the mass of the star. A higher value of entropy also results in the larger radius of the PNS. The astrophysical observational data from PSR J0740+6620 with radius and mass range $M=2.15^{+0.10}_{-0.09}M_\odot$ and PSR J1614-2230 with $M=1.97\pm0.04M_\odot$ has also been shown in Fig. \ref{M-R} \cite{Demorest_2010, Cromartie_2019}. We observe that the G2$^{*}$ parameter set does not satisfy the observational data limit of PSR J0740+6620 for the maximum mass at $S = 0$ and $1$ and the NICER data constraint of the radius for the canonical star ($1.4 M_\odot$) \cite{miller2021radius, pang2021nuclearphysics}. Simultaneously, the maximum mass predicted by the IOPB-I parameter set occur outside the range of PSR J1614-2230. The central temperature $T_{Ce}$ of the star also rises with an increase in the star's mass.
	\begin{table}
		\centering
		\caption{Maximum mass and radius of the PNS for G2$^{*}$ and IOPB-I parameter sets calculated using constant entropy EoS for S = 0, 1 and 2. $T_{Ce}$ denotes the central temperature of the PNS. }
		\scalebox{1.1}{
			\begin{tabular}{|l|lll|lll|}
				\hline
				\multirow{2}{*}{\begin{tabular}[c]{@{}l@{}}Entropy\\  ($S$)\end{tabular}} & \multicolumn{3}{l|}{G2$^{*}$} & \multicolumn{3}{l|}{IOPB-I} \\ 
				\cline{2-7} & 
				{\begin{tabular}[c]{@{}l@{}}$M$\\($M_\odot$)\end{tabular}}&
				{\begin{tabular}[c]{@{}l@{}}$R$\\(km)\end{tabular}}&
				{\begin{tabular}[c]{@{}l@{}}$T_{Ce}$\\(MeV)\end{tabular}}&
				{\begin{tabular}[c]{@{}l@{}}$M$\\($M_\odot$)\end{tabular}}&
				{\begin{tabular}[c]{@{}l@{}}$R$\\(km)\end{tabular}}&
				{\begin{tabular}[c]{@{}l@{}}$T_{Ce}$\\(MeV)\end{tabular}}
				\\ \hline
				0&1.993& 11.96&0 & 2.158& 11.95& 0\\ \hline
				1& 2.051& 12.38& 40.68& 2.197& 13.00& 41.23\\ \hline
				2& 2.099& 13.76& 80.02& 2.265& 14.30& 81.35\\ \hline
		\end{tabular}}
		\label{table3}
	\end{table}
	\section{Conclusions}\label{conc}
	We perform a detailed analysis of the thermal and nuclear properties for the hot nuclear matter by adopting the temperature-dependent relativistic mean-field formalism for the well-known G2$^{*}$ and recently developed IOPB-I parameter sets. We explore the effects of temperature on the binding energy, pressure, entropy, thermal index, symmetry energy, and slope parameter for symmetric nuclear matter. We observe that the strength of confinement of the nuclear matter decreases with an increase in temperature, and the pressure density increases considerably. The critical temperature for the liquid-gas phase transition in an asymmetric nuclear matter system has also been calculated and collated with the experimentally available data. The crucial temperature for G2$^{*}$ and IOPB-I parameter sets are 14.10 and 13.82 MeV, respectively. We determine that the thermal index of the system shows a maximum around the saturation density for both the parameter sets. The relativistic and non-relativistic behavior of nuclear matter has been discussed in the wake of the thermal index. The simulation for the estimation of neutrino emissivity through the direct Urca process of the supernovae remnants has also been performed, which manifests some exciting results about the thermal stabilization, in-homogeneous phases of matter at sub-nuclear densities, and the evolution of the proto-neutron star. We observe that the neutrino emission is responsible for the cooling only in the early evolutionary cycle of the star and the fragmentary excerpt of the supernovae remnants calms down more rapidly through the neutrino emission. We also analyze the mass-radius profile of the proto-neutron star obtained with the constant entropy approach and found that the maximum mass value and canonical star radius for G2$^{*}$ parameter set does not satisfy the constraints of the pulsars' observational data and NICER data respectively.
	\section*{Acknowledgements}
	This work was partially supported by FOSTECT Project Code: FOSTECT.2019B.04, and FAPESP Project Nos. 2017/05660-0.
	\bibliographystyle{apsrev4-1}
	\bibliography{ft.bib}
\end{document}